\documentclass[12pt]{amsart}
\usepackage{amsfonts, amsmath, amssymb}
\usepackage{amsthm}
\usepackage{graphicx}
\usepackage{float}
\usepackage{enumitem}

\usepackage[utf8]{inputenc} 
\usepackage{enumitem}
\usepackage{lipsum}  

\usepackage{graphicx}
\usepackage{subcaption}
\usepackage[utf8]{inputenc}
\usepackage{url}

\numberwithin{Theorem}{section}

\theoremstyle{definition}

\theoremstyle{remark}

\numberwithin{equation}{section}
\expandafter\chardef\csname pre amssym.def
at\endcsname=\the\catcode`\@ \catcode`\@=11
\def\undefine#1{\let#1\undefined}
\def\newsymbol#1#2#3#4#5{\let\next@\relax
 \ifnum#2=\@ne\let\next@\msafam@\else
 \ifnum#2=\tw@\let\next@\msbfam@\fi\fi
 \mathchardef#1="#3\next@#4#5}
\def\mathhexbox@#1#2#3{\relax
 \ifmmode\mathpalette{}{\m@th\mathchar"#1#2#3}%
 \else\leavevmode\hbox{$\m@th\mathchar"#1#2#3$}\fi}
\def\hexnumber@#1{\ifcase#1 0\or 1\or 2\or 3\or 4\or 5\or 6\or 7\or 8\or
 9\or A\or B\or C\or D\or E\or F\fi}

\font\teneufm=eufm10 \font\seveneufm=eufm7 \font\fiveeufm=eufm5
\newfam\eufmfam
\textfont\eufmfam=\teneufm \scriptfont\eufmfam=\seveneufm
\scriptscriptfont\eufmfam=\fiveeufm

\catcode`\@=\csname pre amssym.def at\endcsname

\begin{document}





\title[]{Optimal quarantine strategies for the COVID-19 pandemic in a population with a discrete age structure}

\author[Gondim, Machado]{João A. M. Gondim$^1$, Larissa Machado$^2$}

\email{joao.gondim@ufrpe.br}

\bigskip

\maketitle

\centerline {$^1 \, $Unidade Acadêmica do Cabo de Santo Agostinho} \par \centerline{
Universidade Federal Rural de Pernambuco} \par  \centerline{Cabo de Santo Agostinho, PE
CEP 54518-430 Brazil} \vspace{0.2cm}

\centerline {$^2 \, $Departamento de Matem\'atica} \par \centerline{
Universidade Federal de Pernambuco} \par  \centerline{Recife, PE
CEP 50740-540 Brazil} \vspace{0.2cm}

\begin{abstract}
The goal of this work is to study the optimal controls for the COVID-19 epidemic in Brazil. We consider an age-structured SEIRQ model with quarantine compartment, where the controls are the quarantine entrance parameters. We then compare the optimal controls for different quarantine lengths and distribution of the total control cost by assessing their respective reductions in deaths in comparison to the same period without quarantine. The best strategy provides a calendar of when to relax the isolation measures for each age group. Finally, we analyse how a delay in the beginning of the quarantine affects this calendar by changing the initial conditions.
\end{abstract}
\centerline{{\bf AMS :}
92D30, 93C15, 34H05, 49K15}
\vspace{0.4cm}
\centerline{ {\bf Key Words:} 
COVID-19, Quarantine, SEIR model, Optimal control} \vspace{0.4cm} 

\newpage 

\section{Introduction}

At the end of 2019 a novel coronavirus emerged in the city of Wuhan, China. In January 2020 the disease was given the name COVID-19 and, by mid February, China already faced over 60 thousand cases \cite{worldometers}. Many scientists then began to model disease to forecast its worldwide impact (\cite{wu2020nowcasting}, \cite{ferguson2020report}), even influencing policies of many governments.

As the disease spread across Europe and the United States, some countries were forced to implement quarantines or even lockdowns to mitigate the harms and were able to do the so called ``flattening of the curve'', i.e., to postpone and dampen the maximum number of active cases. By mid May, Brazil is entering this stage, with some states declaring stricter quarantine policies.

Optimal control theory has been applied to general epidemic models (\cite{behncke2000optimal}, \cite{mateus2018optimal}) as well as to specific disease models such as HIV (\cite{joshi2002optimal}, \cite{fister1998optimizing}, \cite{kirschner1997optimal}), tuberculosis (\cite{silva2013optimal}, \cite{jung2002optimal}) and influenza (\cite{lee2010optimal}). Recently, a few works applying optimal control theory to the COVID-19 pandemic, such as \cite{grigorieva2020optimal} and \cite{djidjou2020optimal}, have appeared. This paper focuses on a SEIR model with quarantine as was proposed in \cite{jia2020modeling}, but dividing the population in age groups as in \cite{castilho2020assessing}. This is particularly important since COVID-19 has worse consequences on the elderly than it does on younger people. 

Our goal is to calculate the optimal quarantine strategies numerically for different choices of parameters in the model, which reflect the decisions governments must make when implementing these policies, such as evaluating the economical costs of the quarantine for each of the age groups and when to start implementing the measures. Then, we compare the controls by looking at how they reduce deaths in comparison to the same period without quarantine. The best strategy gave us a calendar of when to relax the measures in each of the age groups.

\section{The age-structured SEIRQ model}
 
 Our model consists of a classical SEIR model with a quarantined class. Besides, we assume that the population has an age structure (see \cite{thieme2001disease}, \cite{inaba2006mathematical} for models with a continuous age structure and \cite{zhou2004dynamics}, \cite{zhou2019global} for models with a discrete one). There are three age groups, described in Table \ref{idades}.
 
 \begin{table}[!h]
 	\centering
 	\caption{Description of the age groups.}
 	\label{idades}
 	\begin{tabular}{cc}
 		\hline  
 		\\
 		Age group 	  &  Description    \\ 
 		\\ \hline
 		\\

 		1 	  &  Young people, aged 0 to 19  \\ 
 		\\
 		2   &	Adults, aged 20 to 59 \\ 
 		\\ 
 		3     &  Elderly, aged 60 onwards. \\ 
 		\\
 		\hline
 	\end{tabular}%
\end{table}

Let $S_i(t)$, $E_i(t)$, $I_i(t)$, $R_i(t)$ and $Q_i(t)$ be the number of susceptible, exposed, infected, recovered and quarantined individuals in each age group at time $t \geq 0$, respectively. We assume that the total population $$N(t) = \sum_{i=1}^3 \left( S_i(t) + E_i(t) + I_i(t) + R_i(t) + Q_i(t)\right)$$ is constant since we are only dealing with a short time frame in comparison to the demographic time scale. The equations, for $i \in \{1,2,3\}$, are as follows
 
 \begin{equation}
 \begin{aligned}
     S_i'(t) &= -\frac{S_i(t)}{N(t)}\left( \sum_{j=1}^3 \beta_{ij}I_j(t)\right) - u_i(t)S_i(t)+\lambda Q_i(t) \ , 
     \\[0.4cm]
     E_i'(t) &= \frac{S_i(t)}{N(t)}\left( \sum_{j=1}^3 \beta_{ij}I_j(t)\right) - \sigma_i E_i(t) \ ,
     \\[0.4cm]
     I_i'(t) &= \sigma_i E_i(t) - \gamma_i I_i(t) \ ,
     \\[0.4cm]
     R_i'(t) &= \gamma_i I_i(t) \ ,
     \\[0.4cm]
     Q_i'(t) &= u_i(t) S_i(t) - \lambda Q_i(t). \\
 \end{aligned}
 \end{equation}
 
All parameters are nonnegative. $\beta_{ij}$ is the transmission coefficient from age group $i$ to age group $j$. Typically, it will be assumed that $\beta_{ij} = \beta_{ji}$ for all $i,j$. $\sigma_i$ and $\gamma_i$ are the latency and recovery periods, respectively, for age group $i$. $\lambda$ is the exit rate from the quarantine. Our controls are the $u_i(t)$, which denote the percentage of susceptible individuals in each age group that are put into quarantine at time $t$. As such, they satisfy, a priori,
\begin{equation}
    0 \leq u_i(t) \leq 1 \ , \quad i \in \{1,2,3\}.
    \label{percui}
\end{equation}

However, it is unrealistic to expect an entire population to stay under quarantine for a long time. There are essential workers such as healthcare professionals and police officers that cannot stay at home during these times. As most of these workers are in age group 2, we suppose that all of age groups 1 and 3 can be quarantined (for age group 3, indeed, this is especially important since they are a risk group for the COVID-19 pandemic). Thus, we shall loosen \eqref{percui} by considering, for example, $$0 \leq u_2(t) \leq u_{\max}.$$

Evaluating $u_{\max}$ is one of the tasks of each government's authorities. In this paper, we fix this parameter at $u_{\max} = 0.9$. This means that 
\begin{equation}
    0 \leq u_1(t) \leq 1, \ \ 0 \leq u_2(t) \leq 0.9, \ \ 0 \leq u_3(t) \leq 1.
    \label{percui2}
\end{equation}

 Let $$N_i(t) = S_i(t) + E_i(t) + I_i(t) + R_i(t) + Q_i(t)$$ be the total population of age group $i$. Adding the equations in the system above, we see that $N_i(t)$ is also constant for $i \in \{1,2,3\}$. Since $R_i(t)$ only appears in the other equations as a part of $N_i(t)$, we substitute the equations for $R_i'i(t)$ by $N_i'(t)$.. Hence, we may also consider the system of
 
 \begin{equation}
  \begin{aligned}
     S_i'(t) &= -\frac{S_i(t)}{N(t)}\left( \sum_{j=1}^3 \beta_{ij}I_j(t)\right) - u_i(t)S_i(t)+\lambda Q_i(t) \ , \\[0.4cm]
     E_i'(t) &= \frac{S_i(t)}{N(t)}\left( \sum_{j=1}^3 \beta_{ij}I_j(t)\right) - \sigma_i E_i(t) \ ,\\[0.4cm]
     I_i'(t) &= \sigma_i E_i(t) - \gamma_i I_i(t) \ , \\[0.4cm]
     Q_i'(t) &= u_i(t) S_i(t) - \lambda Q_i(t) \ , \\[0.4cm]
     N_i'(t) &= 0.\\
 \end{aligned}
 \label{sistcomN}
 \end{equation}
    
Parameter values are taken from \cite{castilho2020assessing} and are listed in Table \ref{parameters-struc}. The data fitting was performed using an adaptation of a least-squares algorithm from \cite{martcheva2015introduction}.
     
\begin{table}[!h]
 	\centering
 	\caption{Parameter values (data from \cite{castilho2020assessing}).}
 	\label{parameters-struc}
 	\begin{tabular}{cccc}
 		\hline  
 		\\
 		Parameter 	  &  Value   & Parameter   & Value\\ 
 		\\ \hline
 		\\

 		$\beta_{11}$ 	    &  1.76168 & $\sigma_1$    	    &  0.27300 \\ 
 		\\
 		$\beta_{12}$    	    &  0.36475 & $\sigma_2$     &  0.58232 \\ 
 		\\ 
 		$\beta_{13}$     &  1.32468 & $\sigma_3$      	&  0.69339\\ 
 		\\
 		$\beta_{22}$      	&  0.63802 & $\gamma_1$   		 &  0.06862 \\ 
 		\\
 		$\beta_{23}$   		 &  0.35958 & $\gamma_2$   		 &  0.03317 \\ 
 		\\
 		$\beta_{33}$ 	    &  0.57347 & $\gamma_3$   		 &  0.35577\\ 
 		\\\hline
 	\end{tabular}%
\end{table}

To see how the numbers of infections and recoveries are distributed in the three age groups, we refer to the data available in \cite{esp}, shown in Table \ref{totaiscasos}. For simplification, we suppose that the difference between the number of cases and the number of deaths represents the number of recoveries. This is not necessarily correct, because some of the patients that we considered as recovered might still carry the disease, but we use this approach due to the scarcity of information regarding recoveries we currently have. The respective distributions are shown in Table \ref{distributions}.

 \begin{table}[!h]
 	\centering
 	\caption{Number of cases, deaths and recoveries by age group (\cite{esp}).}
 	\label{totaiscasos}
 	\begin{tabular}{cccc}
 		\hline  
 		\\
 		Age group 	  &  Cases &  Deaths & Recoveries    \\ 
 		\\ \hline
 		\\

 		1 	    &  2448 & 7 & 2441  \\ 
 		\\
 		2   	&  113059 & 891 & 112168 \\ 
 		\\ 
 		3     &  121928 & 17948 & 103980\\ 
 		\\
 		Total & 237435 & 18846 & 218589 \\
 		\\
 		\hline
 	\end{tabular}%
\end{table}

\begin{table}[!h]
 	\centering
 	\caption{Distribution of infections and recoveries by age group.}
 	\label{distributions}
 	\begin{tabular}{ccc}
 		\hline  
 		\\
 		Age group 	  &  \% of cases &  \% of recoveries    \\ 
 		\\ \hline
 		\\

 		1 	    & 1.03\% & 1.12\%  \\ 
 		\\
 		2   	&  47.62\% & 51.31\%  \\ 
 		\\ 
 		3     &  51.35\% & 47.57\% \\ 
 		\\
 		Total & 100\% & 100\%  \\
 		\\
 		\hline
 	\end{tabular}%
\end{table}

 According to \cite{worldometers}, Brazil had as of May 13, 2020 a total of 97,575 active COVID-19 cases. Even though there seems to be a large subnotification in the country \cite{russel}, this number will be considered as the total number of infected individuals nonetheless. To estimate the number of exposed cases, we look at data from May 8, 2020, since the mean incubation period of the disease is thought to be around 5 days \cite{lauer2020incubation}. Once again according to \cite{worldometers}, at this time Brazil had 76,603 active cases, so this gives us an estimation of 20,972 exposed cases. We also suppose that these cases follow the age distributions of cases from Table \ref{distributions}. Finally, as of May 8, there were 65,124 recovered cases in Brazil \cite{worldometers}. 
 
 Therefore, our initial time will consist of data from Brazil as of May 8, 2020. The total population is assumed to be 200 million, divided into 40\% young people, 50\% adults and 10\% elderly. We also assume that there are no quarantined individuals when the simulation starts. Since the numbers of exposed, infected and recovered are very small in comparison to the total population, we assume that the initial number of susceptible individuals is equal to the total population of the respective age group. The initial conditions of all variables, rounded to the nearest integers, are listed in Table \ref{condini}.
 
 \begin{table}[!h]
 	\centering
 	\caption{Initial conditions.}
 	\label{condini}
 	\begin{tabular}{cccc}
 		\hline  
 		\\
 		Class 	  &   $i = 1$ &  $i = 2$ & $i = 3$    \\ 
 		\\ \hline
 		\\

 		Susceptible & 80 million  & 100 million & 20 million  \\ 
 		\\
 		Exposed   	&  216 & 9987 & 10769  \\ 
 		\\ 
 		Infected     &  789 & 36478 & 39335 \\ 
 		\\
 		Recovered & 729 & 33415 & 30979  \\
 		\\
 		Quarantined & 0 & 0 & 0 \\
 		\\ \hline
 	\end{tabular}%
\end{table}

 \section{The optimization problem}
 
 Using system \eqref{sistcomN}, we consider the functional to be minimized as 
 \begin{equation}
     J = \int_0^T \sum_{i=1}^3 \left(I_i(t) + B_iu_i^2(t)\right)dt
 \end{equation}
 
  In the formula above, $T$ is the quarantine duration and the parameters $B_i$ are the costs of the control. We assume that $B_i > 0$ for $i \in \{1,2,3\}$ and that
  \begin{equation}
      B_1+B_2+B_3 = B,
      \label{totalcontrol}
  \end{equation}
  where $B \in \mathbb{R}$ is the total control cost. Sufficient conditions for the existence of the optimal controls follow from standard results from optimal control theory. For instance, we can use Theorem 2.1 in \cite{joshi2006optimal} to show that the optimal control exists. Pontryagin's maximum principle (\cite{pontryagin2018mathematical}, \cite{lenhart2007optimal}) establish that optimal controls are solutions of the Hamiltonian system with Hamiltonian function
 \begin{equation}
 \begin{aligned}
     H &= \sum_{i=1}^3 \left(I_i(t)+B_i u_i^2(t)\right)\\
     &+ \sum_{i=1}^3 \left(\lambda_i^S S_i'(t)+\lambda_i^E E_i'(t)+\lambda_i^I I_i'(t) + \lambda_i^Q Q_i'(t) + \lambda_i^N N_i'(t)\right), \\
\end{aligned}
\end{equation}
where $\lambda_i^S$, $\lambda_i^E$, $\lambda_i^I$, $\lambda_i^Q$ and $\lambda_i^N$ are the adjoint variables. These variables must satisfy the adjoint equations 
\begin{equation}
\lambda_i^{C \prime}= - \frac{\partial H}{\partial C_i}, 
\end{equation}
%
%
where $i \in \{1,2,3\}$ and $C \in \{S,E,I,Q,N\}$. The adjoint system is detailed in \eqref{sistadjunt} below.
\begin{equation}
    \begin{aligned}
    \lambda_i^{S \prime} &= \frac{1}{N}\left( \sum_{j=1}^3 \beta_{ij}I_j\right) \left( \lambda_i^S - \lambda_i^E\right) + u_i(t)\left( \lambda_i^S - \lambda_i^Q \right) \ ,\\[0.4cm]
    \lambda_i^{E \prime} &= \sigma_i\left(\lambda_i^E - \lambda_i^I\right) \ , \\[0.4cm]
    \lambda_i^{I \prime} &= -1 + \frac{1}{N} \sum_{j=1}^3 \beta_{ji} S_j \left(\lambda_j^S-\lambda_j^E\right) + \gamma_i \lambda_i^I \ , \\[0.4cm]
    \lambda_i^{Q \prime} &= \lambda \left(\lambda_i^Q-\lambda_i^S\right) \ , \\[0.4cm]
    \lambda_i^{N \prime} &= \frac{1}{N^2} \sum_{k=1}^3\sum_{j=1}^3 \beta_{kj}S_kI_j\left(\lambda_k^E-\lambda_k^S\right). \\
    \end{aligned}
    \label{sistadjunt}
\end{equation}
 
 The adjoint variables also must satisfy the transversality conditions 
 
 \begin{equation}
     \lambda_I^S(T) = \lambda_i^E(T) = \lambda_i^I(T) = \lambda_i^Q(T) = \lambda_i^N(T) = 0,
 \end{equation}
 for $i \in \{1,2,3\}$.
 
 Finally, the optimality conditions come from solving 
 \begin{equation}
 \frac{\partial H}{\partial u_i} = 0.
 \end{equation}
 This results in 
 \begin{equation}
     u_i^* = \frac{\left(\lambda_i^S - \lambda_i^Q\right)S_i}{2B_i}.
     \label{condoptm}
 \end{equation}
 
 Since we are considering bounded controls (because of \eqref{percui2}), the $u_i^*$ are calculated using
 \begin{equation}
     u_i^* = \min\left\{u^i_{\max},\max\left\{0,\frac{\left(\lambda_i^S - \lambda_i^Q\right)S_i}{2B_i}\right\}\right\}, \\
     \label{boundedcontrol}
 \end{equation}
where $u^1_{\max} = u^3_{\max} = 1$ and $u^2_{max} = 0.9$.

Uniqueness of the optimal controls (at least for small enough $T$) also follow from standard results, such as Theorem 2.3 in \cite{joshi2006optimal}. Numerical solutions of systems \eqref{sistcomN} and \eqref{sistadjunt} can be found by a forward-backward sweep method \cite{lenhart2007optimal}. The algorithm starts with an initial guess of the controls $u_1$, $u_2$ and $u_3$ and then solves \eqref{sistcomN} forward in time. After this first part, it uses the results and the initial guesses to solve \eqref{sistadjunt} backward in time and new controls are defined following \eqref{boundedcontrol}. This process continues until it converges.

\section{Comparison of optimal controls for different control costs}
 
 Quarantines are not just a matter of public health, for they also present economic questions, for example. This means that we must pay close attention to the control costs $B_i$. These numbers reflect how the population is capable of dealing with the quarantine of the respective age group. Smaller values of $B_i$ mean that the population can withstand a stricter quarantine without many economical side effects. This is not possible, on the other hand, for bigger values of $B_i$. 
 
 Since the bigger economic toll of the quarantine lies on the adults (because they form almost all of the economic active population), we assume that $B_2$ is the greatest of the three values. As the isolation of the young implies closing schools, this educational impact makes $B_1$ the second highest cost, albeit much smaller than $B_2$.

How the total cost $B$ is distributed among the age groups determines the shape of the optimal controls. To study this relationship, we let $B = 5000$ and define four cost distributions $D_1$, $D_2$, $D_3$ and $D_4$, detailed in Table \ref{controldist}.

\begin{table}[!h]
 	\centering
 	\caption{The four distributions considered for the total control $B$.}
 	\label{controldist}
 	\begin{tabular}{cc}
 		\hline  
 		\\
 		Label	  &   Distribution    \\ 
 		\\ \hline
 		\\

 		$D_1$ &  $B_1 = 490$, $B_2 = 4500$, $B_3 = 10$. \\ 
 		\\
 		$D_2$ &  $B_1 = 980$, $B_2 = 4000$, $B_3 = 20$.  \\ 
 		\\ 
 		$D_3$ &  $B_1 = 385$, $B_2 = 4600$, $B_3 = 15$.  \\ 
 		\\ 
 		$D_4$ & $B_1 = 1485$, $B_2 = 3500$, $B_3 = 15$. \\
 		\\ \hline
 	\end{tabular}%
\end{table}

\begin{figure} 
	\centering
	\includegraphics[scale=0.75,trim={3.5cm 3.5cm 3.5cm 3.0cm}]{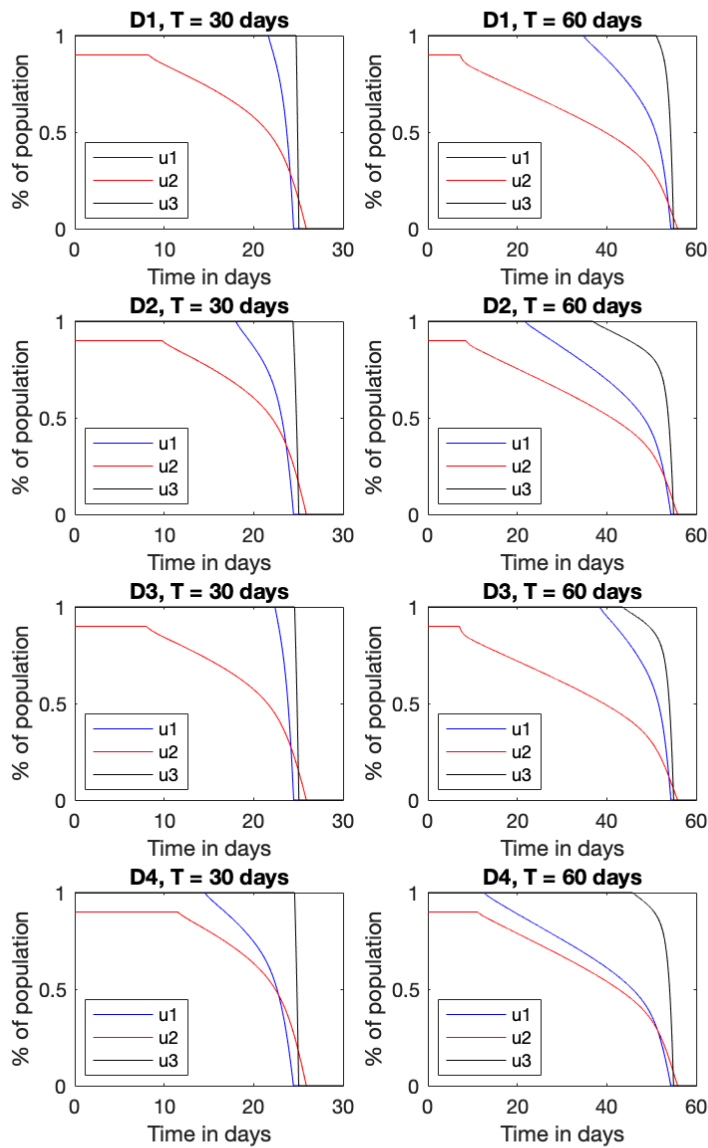}
	\caption{The optimal controls for the four cost distributions and different quarantine durations.}
	\label{fig:controls1} 
\end{figure}

Figure \ref{fig:controls1} shows plots of the controls for $T = 30$ and $T = 60$ days. The overall behavior is that the quarantine is relaxed firstly for the adults, then for the young and lastly for the elderly. The differences lie in the times it takes for these relaxations, which are described in Table \ref{timerelax}.

\begin{table}[!h]
 	\centering
 	\caption{How long it takes until quarantine relaxation.}
 	\label{timerelax}
 	\begin{tabular}{cccc}
 		\hline  
 		\\
 		Distribution & Age group  &   $T = 30$ & $T = 60$    \\ 
 		\\ \hline
 		\\

 		  &      1 &  22 days & 35 days \\ 
 		\\
 		$D_1$ &  2 &  9 days  & 7 days  \\ 
 		\\ 
 		  &      3 &  25 days & 52 days  \\ 
 		\\ \hline
 		\\
 		  &      1 &  18 days & 22 days \\ 
 		\\
 		$D_2$ &  2 &  11 days  & 10 days  \\ 
 		\\ 
 		  &      3 &  25 days & 39 days  \\ 
 		\\ \hline
 		\\
 		  &      1 &  23 days & 40 days \\ 
 		\\
 		$D_3$ &  2 &  9 days  & 7 days  \\ 
 		\\ 
 		  &      3 &  25 days & 43 days  \\ 
 		\\ \hline
 		\\
 		  &      1 &  15 days & 13 days \\ 
 		\\
 		$D_4$ &  2 &  12 days  & 11 days  \\ 
 		\\ 
 		  &      3 &  25 days & 47 days  \\ 
 		\\ \hline
 	\end{tabular}%
\end{table}

Our goal now is to compare these distributions by investigating the number of deaths caused by the pandemic at the end of the quarantine for each one of them. As in \cite{castilho2020assessing}, the deaths will be calculated as a fraction of the recovered, since there is no disease induced mortality in our model. From Table \ref{totaiscasos}, we can derive the case fatality rates $\mu_1$, $\mu_2$ and $\mu_3$ for age groups 1, 2 and 3, respectively. The results are
\begin{equation}
    \mu_1 = \frac{7}{2448} = 0.003, \ \ \mu_2 = \frac{891}{113059} = 0.008, \ \ \mu_3 = \frac{17948}{121928} = 0.147.
    \label{casefatalities}
\end{equation}

Let $\mathcal{D}$ denote the number of deaths at the end of the quarantine due to the disease. By our discussion above, we can write
\begin{equation}
    \mathcal{D} = \mu_1 R_1(T) + \mu_2 R_2(T) + \mu_3 R_3(T).
    \label{mortes}
\end{equation}

Because of the uncertain nature of the parameters and due to the high number of unreported cases, we do not show the crude numbers of $\mathcal{D}$ for the four distributions. The approach we use is to select the smallest values as unit and then scale the other values accordingly. The results are displayed in Table \ref{mortesprop}. We also included $D_0$, representing the outcome of no quarantine.

\begin{table}[!h]
 	\centering
 	\caption{Proportion of deaths in comparison to the lowest outcomes.}
 	\label{mortesprop}
 	\begin{tabular}{ccc}
 		\hline  
 		\\
 	        &   $T = 30$ & $T = 60$   \\ 
 		\\ \hline
 		\\
        $D_0$ &  110 & 155 \\ 
        \\
 		$D_1$ &  1.0025 & 1.0025 \\ 
 		\\
 		$D_2$ &  1.0015 & 1.0013  \\ 
 		\\ 
 		$D_3$ &  1.0050 & 1.0029  \\ 
 		\\ 
 		$D_4$ & 1 & 1 \\
 		\\ \hline
 	\end{tabular}%
\end{table}

On the one hand, we see that all the optimal controls reduces the number of deaths in more than 100 times in comparison to ``doing nothing'' scenario. On the other hand, the four distributions produce very similar results, with $D_4$ being slightly better for both quarantine lengths.

\begin{figure} 
	\centering
	\includegraphics[scale=0.7,trim={3.5cm 10cm 3.5cm 10.5cm}]{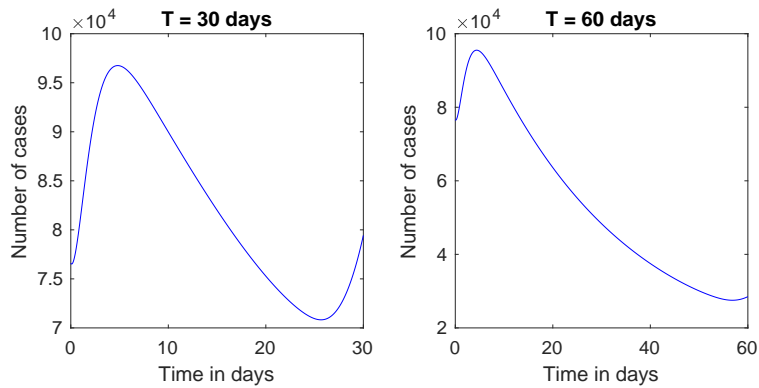}
	\caption{Curves of infections for $D_4$.}
	\label{fig:infectados1} 
\end{figure}

 
For a better visualisation, we plot the graphs of the total number of infections for distribution $D_4$ in Figure \ref{fig:infectados1}.

Both curves reach their minimum around the same time that the controls reach zero. For a quarantine of 60 days, the number of cases at the end is, indeed, much smaller than they are in the beginning. However, for a shorter quarantine, there could even be more cases than the initial total.

This fact shows that quarantines cannot be too short, or else the overall situation in the end could be worse than in the beginning. Moreover, the number of cases start to go up again towards the end. 

\begin{figure}[ht]
	\centering
	\includegraphics[scale=0.65,trim={3.5cm 10cm 3.5cm 10.5cm}]{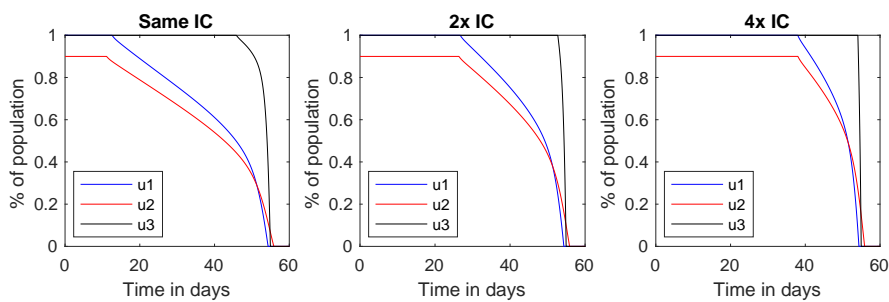}
	\caption{Plots of the optimal controls for different initial conditions.}
	\label{fig:controls4} 
\end{figure}

To finish this Section, we analyse how the initial conditions affect the optimal control. We can interpret this as a way to see what happens if it takes too long for these measures to be implemented. We do this by considering initial conditions of exposed, infected and recovered twice and four times as much as their original values. As of May 13, 2020, the number of active cases in Brazil doubles every 10 days \cite{worldometers}, so this means waiting 10 or 20 days, respectively, to start the quarantine. In the plots of Figure \ref{fig:controls4}, we consider distribution $D_4$ and quarantine length of 60 days.

The plot on the left of Figure \ref{fig:controls4} is the same as the last one in \ref{fig:controls1}. The quarantine is relaxed for the young, adults and elderly after 13, 11 and 47 days, respectively. The plot in the middle has the initial conditions of exposed, infected and recovered multiplied by two. Now, the quarantine is relaxed only after 27, 26 and 53 days, respectively. The last plot multiplies the initial conditions by four. The relaxations happen, respectively, after 39, 38 and 54 days. 

This shows that the quarantines have to be much stricter if there is a delay in their implementation. Moreover, applying \eqref{mortes} with and without quarantine, we see that the deaths are reduced by  factors of 79 and 40 when initial conditions are multiplied by 2 and 4, respectively. With an earlier quarantine entrance, Table \ref{mortesprop} showed that the reduction was by a factor of 155 for distribution $D_4$, so the longer it takes for the quarantine to start, the less effective it is.



 \section{Conclusions}
 
 In this paper we considered an age-structured SEIRQ model, where the quarantine entrance parameters are thought as controls of the system, and we looked for the optimal controls via Pontryagin's maximum principle. After writing down the optimality system, we calculated the optimal controls numerically and analysed how some of the parameters influence the results.
 
 These parameters represent the difficult choices authorities must make, such as deciding how many essential workers are allowed to remain circulating, estimating the economic impact of the quarantine and even when to start it. As such choices are made, the optimal controls give guidelines of how to proceed.
 
 In Section 4, we considered a constant total control cost and distributed it among the age groups in four ways. The distribution with the best results with regard to deaths during the quarantine gave us a calendar of when to relax the isolation measures in the three age groups (for quarantine lengths of 30 and 60 days, respectively):
 \begin{itemize}
     \item For the youngs, the date of relaxation was the 15th or the 13th day.
     \item For the adults, the relaxation started at the 12th day or the 11th day.
     \item For the elderly, it started at the 25th day or the 47th day.
 \end{itemize}
 
The optimal controls that induce this calendar produced a reduction in the number of deaths of 110 and 155 times, respectively, in comparison to the same period of time without quarantine. However, in both cases the number of infected cases reached a minimum just before the end of the simulation, so by the time the quarantine ended, the cases were going up again, even becoming bigger than the original values for the shorter length we considered. This shows that the quarantines are not effective if they are not long enough.
 
 We also showed that waiting too long to start the quarantine makes the period before the relaxation become longer. This also produced a loss in efficacy, since the reduction of deaths due to the quarantine (in comparison to the ``doing nothing'' scenario) appeared to be approximately inversely proportional to the number of initial cases.
 
 In our model we used data from Brazil as initial conditions. Brazil is a very large country, with many cities at different stages of the pandemic. This means that studies such as this one should be made locally to best suit the characteristics of each city. As the plots of Figure \ref{fig:controls4} suggest, the sooner the quarantine is implemented, the shortest the time the controls need to stay at their maximum is. 
 
\section*{Acknowledgement} The authors would like to thank César Castilho (UFPE) for all the valuable discussions and suggestions during the preparation of this manuscript.

\bibliographystyle{amsplain}
\bibliography{references}

\end{document}